# Gaussian Mixture Model-Based Focused Refinement for Enhanced Flexible Structure Determination in CryoEM and CryoET

Muyuan Chen[1*]

[1] Division of CryoEM and Bioimaging, SSRL, SLAC National Accelerator Laboratory, Stanford University, Menlo Park, CA 94025, USA
* Correspondence to: muyuanc@stanford.edu

## Abstract

Dynamic conformational changes of proteins are crucial for their cellular functions. Here we present a unified refinement pipeline for flexible protein structures in both CryoEM and *in situ* CryoET. Using a Gaussian mixture model-based focused alignment procedure, we improve resolution of small domains in highly dynamic proteins and reveal intricate conformational changes. The method corrects the per-subunit motion of TRPV1 and captures the rotary dynamics of ATP synthase within mitochondria.

## Main

Protein structures and their dynamic conformational changes within cells are vital connections between genomic data and functional phenotypes in biological systems. Single particle cryogenic electron microscopy (CryoEM) captures snapshots of many copies of the same protein, allowing for the alignment of 2D particles and reconstruction of high-resolution 3D structures. Analyzing variability among these particles provides insights into the structural dynamic of the macromolecular system[1–3]. *In situ* cryogenic electron tomography (CryoET) offers the ability to visualize protein complexes within living cells and holds promise for studying protein dynamics in their native environments[4–6].

The main method of protein heterogeneity analysis has been multi-reference classification, which groups particles based on structural differences[7]. For proteins exhibiting continuous conformational changes, focused refinement is often employed to address the local flexibility[8,9]. Since the method subtracts of parts of protein density from particles using existing references for alignment of the remaining domains, it requires that the target region is sufficiently large to ensure a detectable signal in noisy particles and that the remaining protein density is closely aligned with the reference to avoid artifacts. Recently, various machine learning approaches have emerged for studying protein structural dynamics from CryoEM data, yielding impressive results[10–13]. However, challenges persist in translating deep neural network (DNN) learned movements into high-resolution structural information and validating conformational changes produced by DNNs.

Previously, we introduced a CryoEM refinement pipeline utilizing Gaussian mixture model (GMM) based protein representations and DNN-based manifold embedding[14,15]. Building on this foundation, we now present a unified refinement pipeline applicable to both single particle CryoEM and *in situ* CryoET. This method enables focused alignment on small domains within highly dynamic protein systems without introducing artifacts, producing structures with improved resolution and resolvability at flexible domains while revealing more detailed conformational changes within the system.

The key concept of this method is to modify the GMM reference during alignment rather than altering the Euler angle assignments of the particles. To align a target domain, we apply a 3D rotation matrix to the coordinates of the Gaussian functions within the domain and then project the modified reference at each particle's orientation as determined by global refinement (Figure S1, left). The entire process is implemented in a differentiable framework using JAX[16], so we can calculate the optimal domain rotation that maximize particle-projection similarity using automatically generated gradients. Multiplying the resulting rotation matrix with the original projection matrix yields new Euler angles, and reconstructing particles using these angles leads to improved structures in the focused region.

A key challenge in focused refinement arises when the target domain is small and its movement range is significant, often lacking sufficient signal for precise alignment. To address this, we utilize a combination of two optimization methods. The first is a direct alignment that optimizes translation and rotation of the target domain via gradient descent. The second method mirrors the heterogeneity analysis routine, using a pair of DNNs[17]: one encoder mapping particles to a lower-dimensional latent space and a decoder outputting the rotation matrix for each particle's focused region.

These two optimization techniques target distinct aspects of structural flexibility (Figure S2). In simulated datasets, if the target region experiences only small random rotation/translation relative to the rest of the protein, direct alignment performs more effectively. Conversely, if the region undergoes long-range movement along a defined trajectory, the DNNs demonstrate a clear advantage, as they can learn the trajectory using collective information from all particles and position each particle along the trajectory. In real-world scenarios where the target domain follows a specific trajectory but with minor random perturbations, a DNN-based optimization followed by direct alignment successfully captures both flexibility types. All examples presented below use the combination of the two methods.

In our first example, we applied our technique to the TRPV1 CryoEM dataset (Figure 1)[18]. Beginning with a homogeneous refinement using CryoSPARC that achieved a resolution of 2.8 Å[19], GMM-based global refinement in EMAN2 improved the resolution of the entire complex to Nyquist (2.4 Å). Due to inherent flexibility, the carboxyl-terminal domain (CTD) remained resolved at a lower resolution (2.7 Å). To demonstrate the method on small regions, we expand the c4 symmetry of the complex, and focus the alignment on the CTD of one subunit. The DNN-based optimization followed by direct alignment enhanced the resolution of the CTD to 2.4 Å. The focused refinement produced a map showing well-defined helices and clearly visible side chains. The Q-score of the focused region also improved[20], with some helices achieving resolvability on par with those in the transmembrane domain (TMD).

Accurate pose determination of individual domains also facilitates more comprehensive and interpretable heterogeneity analyses. Using the TRPV1 dataset, GMM-DNN-based heterogeneity analysis[14] focused on one of the CTDs revealed large-scale motion of the entire domain. Because the large scale movement had been corrected by the focused refinement, the same analysis using orientation assignments from the CTD-focused alignment shows the previously unseen, subtle motion between helices (Figure 1E, Video 1-2).

Next, we expanded our method for CryoET to investigate structural dynamics in highly complex protein systems *in situ*. To develop a unified approach for both CryoEM and CryoET, instead of 3D tomograms, we use 2D particles directly extracted from tilt series. A set of 2D particles, or sub-tilt series, can be reconstructed into a subtomogram using alignment from the tilt series, with consensus subtomogram averages created by reconstructing all 2D particles at the determined orientations (Figure S1, right)[21].

During GMM-based focused refinement, the 2D sub-tilt series are processed similarly to standard CryoEM particles, with all 2D particles corresponding to the same 3D protein sharing a common 3D rotation matrix for the target region. The modified GMM reference is projected at the orientation of each 2D particle, and similarity scores for all 2D particles associated with the same 3D particle are averaged to produce a score for the subtomogram. The gradient of the averaged similarity score with respect to the 3D rotation matrix of the target domain is automatically computed, enabling optimization through DNNs followed by direct alignment. By utilizing 2D sub-tilt series, CryoET particles can be aligned via the same pipeline employed for 2D CryoEM particles, with the addition of a reference list mapping 2D to 3D particles.

To showcase the capability of the method for highly dynamic systems *in situ*, we applied it to a CryoET dataset of the green algae C.reinhardtii[22], analyzing ATP synthase dynamics within mitochondria (Figure 2). Using 30,000 ATP synthase monomer particles from a subset of tomograms, we achieved an averaged structure with a global resolution of 5.6 Å from homogeneous refinement. Focused refinement on the F1 head revealed its characteristic rotary movement and improved the resolution of the domain from 8 Å to 6.5 Å, with alpha-helices clearly resolved, indicating that the DNNs effectively learned the rotary motion from the data and corrected the motion through local alignment (Figure S3, Video 3).

Compared to the F1 head, the central stalk's rotation spans the full 360-degree range, posing a greater challenge for focused refinement. Building on existing work[23], we first identified six key states through focused classification and pre-trained the DNNs using these states. Subsequently, focused refinement yielded an averaged structure of the central stalk at 8.5 Å resolution, with DNNs successfully learning to recover the full rotational profile of the domain (Figure 2E, Video 4). A final composite map was generated from the focused refinements of the F1 head, central stalk, and peripheral stalk, achieving a global "gold-standard" resolution of 5.2 Å.

In conclusion, the GMM-based focused refinement pipeline effectively corrects local flexibility in protein complexes, enhancing the resolution and resolvability of averaged structures. Unlike traditional focused refinement approaches[8], the GMM-based method avoids artifacts from density subtraction and improves alignment accuracy for proteins with flexible domains. The incorporation of DNNs allows for the aggregation of information across all particles to estimate movement trajectories, enhancing alignment accuracy for large-scale coordinated domain motions.

Compared to existing DNN-based heterogeneity analysis methodologies[13,14], the rigid body movement constraints simplify the problem space and improve training convergence. By bypassing the inaccuracy in conformation-to-pose conversion[15], the method retains high-resolution information, yielding better-resolved structures. The enhanced resolution from alignment also serves as an internal validation for the DNN-based heterogeneity analyses. Since focused refinement only alters orientation assignments, it can seamlessly integrate with existing heterogeneity analysis methods, enabling visualization of subtle conformational changes within the system after large-scale motion is corrected.

Finally, the ability to analyze CryoET data extends the investigation of more dynamic protein systems within cells, allowing for higher-resolution structure determination of flexible domains. The analysis of CryoET datasets can uncover intriguing protein structural dynamics in living systems, with improved structural resolutions providing deeper insights into the functioning mechanisms of these biological systems.

**Software availability**

All computational tools described here are implemented in EMAN2, a free and open source software for CryoEM/CryoET imaging processing. The code is available at github.com/cryoem/eman2.

**Data accessibility**

All data used in the paper are publicly available through EMPIAR. Structures produced in this paper will be deposited in the EMDB upon paper acceptance.

**Acknowledgements**

This research is supported by NIH grant R01GM150905 and U.S. DOE under BCIS FWP-101359. The initial data processing and modeling of the ATP synthase dataset was performed by Emily Capper during her summer internship at SLAC.

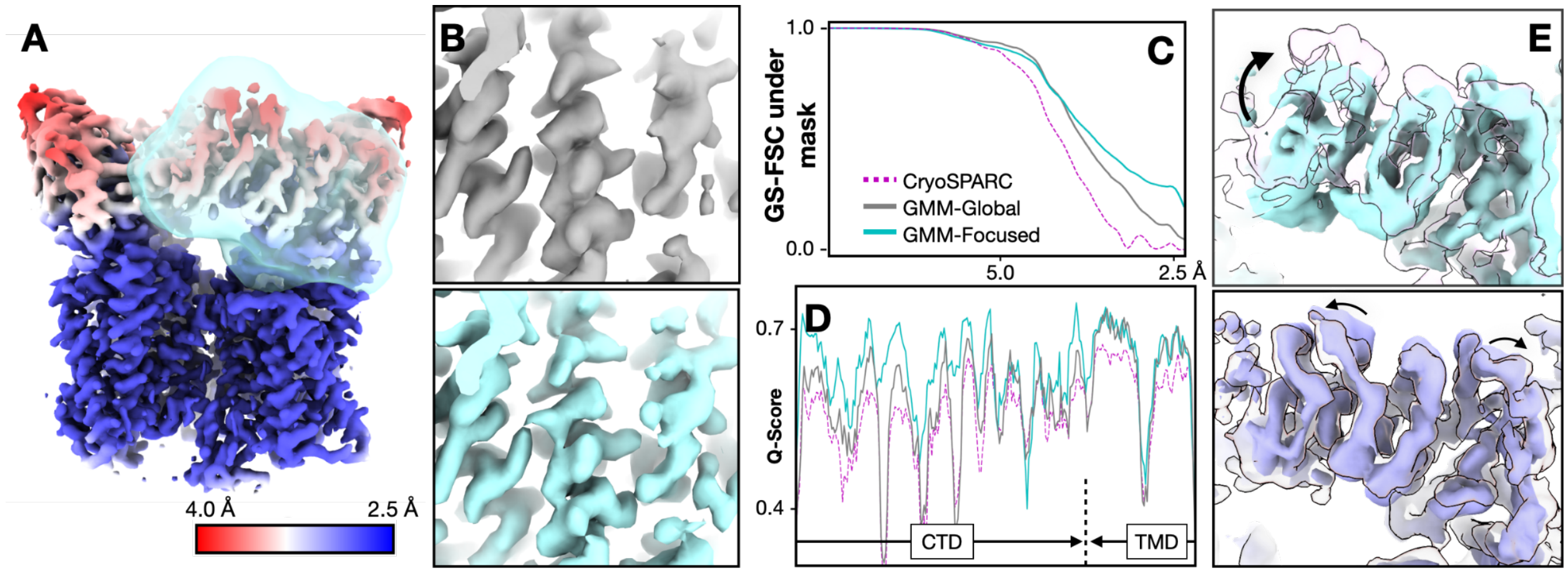

**Figure 1. GMM-based focused refinement on a CryoEM dataset of TRPV1.** (A) Structure of TRPV1 after GMM global refinement, colored by local resolution. Translucent cyan volume marks the target region for focused refinement. (B) Structure at the CTD before (top) and after (bottom) the focused refinement. (C) "Gold-standard" FSC of the target region from CryoSPARC, GMM-based global and focused refinement. (D) Q-score comparison of the three refinements at the CTD inside the focus mask, and TMD outside the mask. (E) Eigen-movement trajectory from GMM-DNN-based heterogeneity analysis focusing on one CTD using initial pose from the global (top) and focused (bottom) refinement. The first frame of the movement is displayed with solid surfaces, and the last frame is displayed with silhouettes.

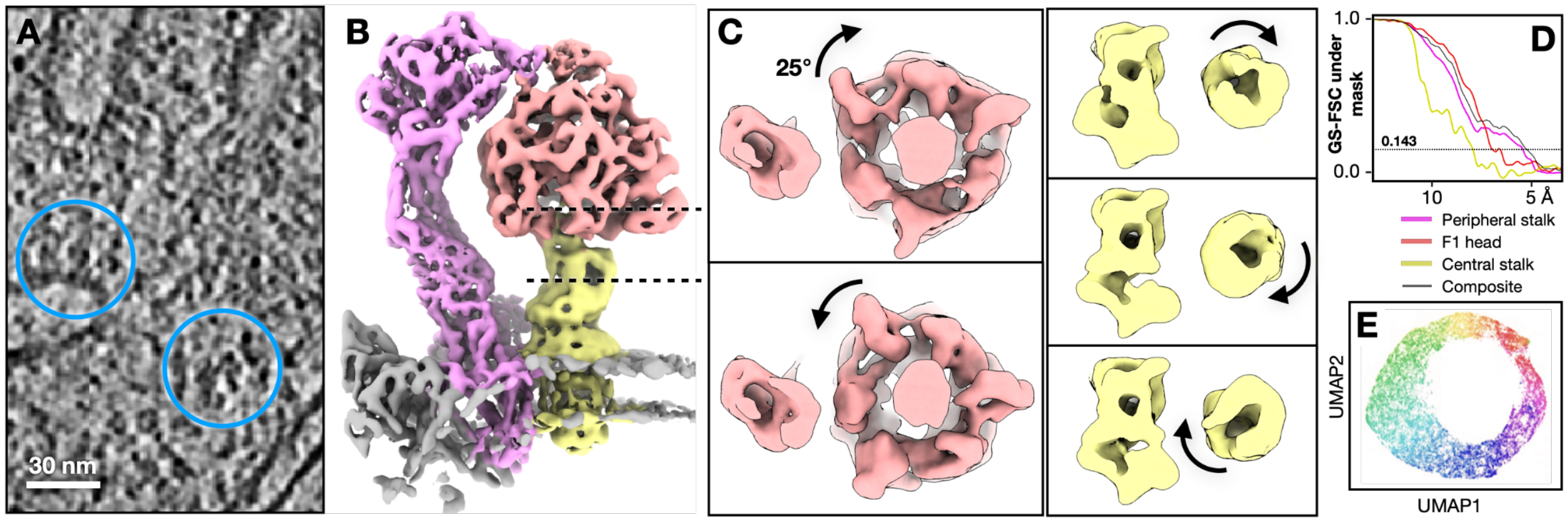

**Figure 2. GMM-based focused refinement on a CryoET dataset of mitochondrial ATP synthase.** (A) A slice view of a mitochondrial tomogram with ATP synthase particles circled. (B) Final composite map of the ATP synthase, colored by the three regions targeted in focused refinement. Dashed lines indicate locations of cross section views in C. (C) Averaged structures of particles at different rotation angles determined by the focused refinement of F1 head (red) and central stalk (yellow). (D) "Gold-standard" FSC of the target region from the focused refinement of the peripheral stalk, F1 head and central stalk. (E) UMAP of the DNN latent space from one of the half-sets in the focused refinement of the central stalk, colored by final rotation angles from direct alignment.

## Methods

### Implementation details

Details of the GMM-based focused refinement workflow can be found in Figure S1. Same as our previous papers[14], a GMM representation of protein density $M$ is defined as:

$$M(\bar{x}) = \sum_{j=1}^{N} A_j e^{-\frac{|\bar{x}-\bar{c}_j|^2}{2\sigma_j^2}}$$

Here, $\bar{c}_j$, $A_j$ and $\sigma_j$ are the coordinates, amplitude and sigma of the $j$th Gaussian function in the GMM. The mask of the target region is defined as a set $S$, and a Gaussian function belongs to the set if the center of the Gaussian in the initial reference structure falls inside the mask. To rotate the structure inside the mask, for particle $i$, a transform matrix $R_i$ is applied to the Gaussian coordinates in the set $S$, and the final GMM becomes:

$$M'(\bar{x}) = \sum_{j \in S} A_j e^{-\frac{|\bar{x}-R_i \bar{c}_j|^2}{2\sigma_j^2}} + \sum_{j \notin S} A_j e^{-\frac{|\bar{x}-\bar{c}_j|^2}{2\sigma_j^2}}$$

The concept of partial rotation of a GMM is similar to our previous work[24], which did not improve structure resolution when used for alignment. Using a different mathematical expression and implemented in a different package, the new version produces correct gradients through automatic differentiation that can align structures properly and retrieve high resolution information.

To calculate $R_i$ given a particle image, we first project the modified GMM $M'$ at $P_i$, the orientation of particle $i$ determined from the global refinement. Similarity between particle and GMM projection is measured by Fourier Ring correlation (FRC), and the gradient $\partial(FRC)/\partial(R_i)$ can be directly computed using automatic differentiation.

To optimize $R$ using DNNs, two three-layer densely connected networks were used. Same as our previous work[14], the encoder takes particle information in the form of per Gaussian gradient with respect to the reference model, $\partial(FRC)/\partial(\bar{c}_j)$, avoiding the impact of particle projection angle. The latent space output has 3 dimensions by default. The decoder takes input from the latent space and outputs $R$ for all particles.

Given optimized $R_i$ for each particle, the new projection matrix is computed by

$$P'_i = R_i \times P_i$$

The new focus-refined map is then generated by reconstructing all particles using their corresponding projection matrices $P'_i$. The updated reference and Euler angle assignment is used for the next iteration of the refinement.

Note that when the focused region covers the entire protein, i.e. $S$ includes all Gaussian functions in the GMM, the focused refinement becomes the same as a global refinement. In practice, when a large enough mask is provided, the focused refinement pipeline produced the exact same results as our previously established GMM-based global refinement[15]. This provides a good sanity check of the algorithm, and also makes it possible to run the GMM-based global refinement pipeline for CryoET data under the same framework.

In the cases where there are multiple flexible domains of a protein, multiple $R_i$ (one transform matrix per target flexible domain) are assigned for each particle. The decoder can output one Euler angle per target domain for every input particle, and the alignment of multiple domains can be optimized simultaneously. After each iteration of alignment, one reconstruction is generated for each target domain from the same particles at different Euler angles. The reconstructions are then merged to form a composite map, using the same method described previously[15], which is used as the reference for the next iteration.

To ensure the refinement process can be validated, the entire global and focused refinement protocol is performed independently on two half-sets of the data. The split of the half-sets only occurs at the very beginning, and the structure obtained from the corresponding half-set in the global refinement is used as the initial reference focused refinement. When DNNs are used for the alignment, one pair of encoder-decoder networks, initialized with independent random variables, is used for each half-set of particles. The only communication between the two half-sets happens at the end of each iteration, when the reconstructions from the two half-sets are aligned to each other and the resolution of the structure is computed by Fourier shell correlation (FSC). Finally, to generate a composite map from the focused refinement of multiple regions of a protein, we first generate one composite map per half-set, then combine the two composite maps to produce the final structure using the same postprocessing procedures used in the homogeneous refinement.

**Focused refinement for CryoET**

For a tomogram reconstructed from $K$ tilted 2D micrographs, $Q_k$, the projection angle of tilt image $k$ with respect to the tomogram is calculated from the tilt series alignment. For a 3D particle in that tomogram, it can be represented as the reconstruction of $K$ 2D particles using the same projection angles $Q_k$. Through subtomogram refinement, the pose of the 3D particle, $T_i$, can be determined by aligning the subtomogram to a reference. The CryoET focused refinement pipeline takes the 2D particles, as well as the tilt series alignment angles $Q_k$ and the subtomogram poses $R_i$, as input. This allows the alignment of CryoEM and CryoET particles under the same framework, and avoids the missing wedge artifacts in subtomograms.

To measure particle-projection similarity for CryoET, the GMM is projected to the orientation of each 2D particle. The projection matrix for the $k$th 2D particle correspond to the $i$th subtomogram is simply computed by:

$$P_{i,k} = Q_k \times T_i$$

FRC is computed for each 2D particle and its corresponding GMM projection. For each subtomogram, the similarity score is defined as the average of the FRC score from all 2D particles it associates with. To perform focused refinement, we only need to optimize the one domain transform $R_i$ for each 3D particle, since the $K$ 2D particles corresponding to the same 3D particle would have the conformation. Same as the single particle CryoEM refinement, the gradient $\partial(Avg(FRC_k))/\partial(R_i)$ can be computed through automatic differentiation.

To train the DNN, the information of each 3D particle is represented as $\partial(Avg(FRC_k))/\partial(\bar{c}_j)$, where $\bar{c}_j$ is the coordinate of the $j$th Gaussian function in the reference GMM. The encoder maps each 3D particle to a point in the latent space, and the decoder outputs the domain transform $R_i$ of that particle. The transform $R_i$ is then applied to the reference GMM, which is then projected at the orientation of 2D particles to compute $Avg(FRC_k)$.

Similar to the CryoEM pipeline, after the optimization of $R_i$, the new projection angle of each 2D particle is:

$$P'_{i,k} = R_i \times P_{i,k}$$

All 2D particles are reconstructed using the new Euler angle to produce the locally aligned 3D averaged structure, which is used as the reference for the next iteration.

**Details of the simulated dataset**

A simulated dataset, generated from the atomic model of a GPCR (PDB:7RMH)[25], was used to show the performance of the optimization methods (Figure S2). Random rotation and translation were applied to the chain corresponding to the transmembrane domain of the protein, while the rest of the structure remains unchanged. A small dataset of 500 particles was simulated for each test case, in which each 2D particle is the projection of an independently perturbed 3D volume. The ground truth projection angles were provided to the refinement, but the random domain perturbation was not. A small amount of noise was added to the particles.

3 types of motion were simulated using this method (Figure S2). First, small random 3D rotation and translation following a normal distribution was applied to the target domain of the reference for each particle. Second, a random movement trajectory of the target domain, including translation and rotation was first selected for all particles, and each particle was assigned to a uniformly distributed random position along that trajectory. Finally, to combine the first two types of movement, in addition to assigning each particle to a random position along the trajectory, a small normally distributed variable was also added to the translation/rotation of the target domain.

We evaluated the results by reconstructing the particles using the assigned Euler angle from the focused alignment, and compared the structure of the domain under mask with the ground truth structure by FSC. For the random motion example, DNN-based optimization produced slightly worse results than direct alignment (3.48 vs 3.55 Å resolution, FSC=0.5 cutoff). In the second example of pre-defined trajectory, the DNN-based optimization outperformed direct alignment (3.46 vs 3.58 Å resolution). Finally, in the case of randomly perturbed trajectory, while direct alignment achieved a better structure than DNN-based optimization initially (3.67 vs 3.94 Å resolution), by running additional rounds of direct alignment on top of the pose output from the DNNs, a structure was produced that has a higher resolution than that from both methods (3.47

Å). Therefore, we designed the pipeline that first uses a DNN to learn the large-scale movement trajectory from all particles, and then correct for the local perturbation using direct alignment.

**Details of the TRPV1 CryoEM dataset**

Micrographs of the dataset were obtained from EMPIAR-10059[18], and the initial processing was performed using CryoSPARC following an existing tutorial. Particles and poses from CryoSPARC homogeneous refinement[19] were imported into EMAN2 for the GMM-based global refinement. From 200k particles, the “gold-standard” resolution of initial refinement was determined to be 2.8Å. The same half-set split from the CryoSPARC homogeneous refinement was kept throughout the GMM-based global and focused refinement, and the same mask was used for FSC calculation.

Three iterations of GMM-based global refinement improved the resolution to 2.4Å, the Nyquist frequency of this dataset. In the averaged structure, the CTD was not as well resolved as the TMD, and the “gold-standard” resolution of the CTD under mask was 2.7 Å. Using a reference initially lowpass filtered to 8Å, three iterations of DNN-based optimization produced an averaged structure with local resolution of 2.9Å. From the pose determined by the DNN-based optimization, two iterations of direct alignment improved the structure further, and the estimated resolution at the CTD was 2.4Å, similar to the resolution of the more rigid TMD. The final map was a composite of the focused refinement map of one CTD and the global refinement map of all remaining regions. The map was sharpened by matching the Fourier power spectrum to the corresponding atomic model, then lowpass filtered to the determined resolution locally. An existing model of TRPV1 was locally fit to the structure using ISOLDE for Q-score calculation[20,26].

To compare how the initial orientation assignment impacts the results of heterogeneity analysis results, we set up the same GMM-DNN based analysis using the previously established method[14]. The heterogeneity analysis used the C4 symmetry expanded particles and focused on the same CTD of the focused refinement. When the heterogeneity analysis uses the initial pose from the global alignment, it revealed the large-scale movement of the CTD, similar to the trajectory determined in the focused refinement and follows a similar movement pattern as reported in previous publications[27]. When the heterogeneity analysis uses the pose from focused refinement, it revealed more subtle movement within the CTD. The slight distortion within the domain has not been observed before, because it is hidden beneath the large-scale domain movement.

**Details of the ATP synthase CryoET dataset**

The CryoET data processing started from motion corrected tilt series obtained from EMPIAR-11830[22]. All initial processing, ranging from tilt series alignment to subtomogram refinement, were performed using the CryoET pipeline in EMAN2[21]. Figure S3 shows the details of the CryoET data processing. Briefly, tilt series were aligned using patch tracking, and the alignment was polished using the gradient based method described previously[28]. The initial model of mitochondrial ATP synthase was generated from 218 handpicked particles, and 25k particles were selected by template matching from 86 tomograms, using the initial model as a reference. After subtomogram refinement with C2 symmetry, we expanded the symmetry and re-extract particles of ATP synthase monomers from the tomograms. 3D classification removed false positive particles, leaving 30k monomer particles for the final voxel and GMM-based global subtomogram refinement (i.e. focused refinement with a large mask), which reached the “gold-standard” resolution of 5.6Å.

The structure of ATP synthase was broken into three pieces for the GMM-based focused refinement, the peripheral stalk, the F1 head and the central stalk (Figure 2B). The peripheral stalk is the most rigid part of the protein, and focused refinement only provided slight improvement in resolution, from 5.4 to 5.2 Å. For the F1 head, the focused refinement improved the “gold-standard” resolution of the domain from 8Å to 6.5Å, and the DNNs used in the alignment recovered the back-and-forth rotary movement of the domain, which covers as much as 25 degrees from the 5th to the 95th percentile of particles.

The focused refinement of the central stalk was more complicated. Because of the full rotation of the domain, its structure in the initial reference from the global refinement has nearly full rotational symmetry. To speed up the convergence of DNN training, we followed a previous work[23], and started from 3D classification focusing on this domain, which revealed six rotatory states similar to what was reported. Although the maps after classification have low resolution (~20Å), they show clear asymmetrical features at the central stalk domain. We took one of the classes as the initial reference for the focused refinement, and rotationally aligned the central stalk of the rest of classes to the reference class, obtaining a rough estimate of the relative orientation of the classes. Then, we pre-trained the DNNs so for each 3D particle input, so that the decoder outputs the relative orientation of the class that the particle was assigned to in the focused classification. Finally, using the classification result as the initial reference and starting from the pre-trained DNNs, we performed the same focused refinement protocol, combining the DNN-based optimization and direct alignment to achieve the final orientation assignment and an averaged structure with 8.5 Å resolution at the target domain.

Finally, composite maps were generated by combining the results from three focused refinement runs, and a molecular model was built by flexibly fitting the AlphaFold structure to the composite map using ISOLDE[26,29].

It is worth noting that because ATP synthase is located on native mitochondrial membrane, the resolution measurement is heavily impacted by masking. All focused refinements performed in the paper use masks with soft falloff to avoid artifacts in the generation of composite maps, and the final resolution is also evaluated using a slightly tighter soft mask that includes some surrounding membrane densities.

**Comparison with existing methods**

The TRPV1 dataset has been experimented using multiple heterogeneity analysis methods by multiple groups, and a comprehensive comparison between packages has been published recently[27]. Notably, none of the methods provides mechanisms to analyze the structural heterogeneity within a local region as small as a single CTD. Among them, 3DFlex was able to improve the resolution of the CTD using the information from the heterogeneity analysis. In their original publication, resolution of the entire CTD was improved from 3.4Å in conventional homogeneous refinement to 3.3Å from local refinement and 3.2Å using 3DFlex[10]. Since the CTD does not always move as a whole, the resolution improvement is limited when they are not treated as individual subunits.

In the same TRPV1 dataset, our previous focused refinement method[15] with sigma-based masking has only slight improvement on the CTD, from 2.7Å to 2.6Å resolution (compared to 2.4Å from our new pipeline). This is likely because the sigma-based masking still uses low resolution information from regions outside the mask, and expect those regions to be less dynamic. In this case, since we only focus on one of four CTDs, the other three CTDs outside the focusing region of the same particle would have the same level of dynamics, making the alignment more challenging.

For the CryoET example of mitochondrial ATP synthase, in the original publication[22], with 88k particles from 261 tomograms, using a combined PyTom-Relion-M data processing pipeline[30–32], including co-refinement with ribosomes, the peripheral stalk of mitochondrial ATP synthase was determined at 5.2Å resolution (see Figure S9 of the paper[22] for a comparison FSC curves). Through 3D classification and focused refinement, an 8.6Å resolution structure of the F1 head was obtained, suggesting that our focused refinement pipeline leads to ~2Å of resolution improvement in flexible regions. The structure and resolution of the full complex was not reported.

Note while the pixel size of EMPIAR-11830 is later calibrated to be 1.91Å[5], we still use the original reported pixel size of 1.96Å to report resolution, so it is comparable with the previous publication. A pixel size of 1.91Å was used to build models.

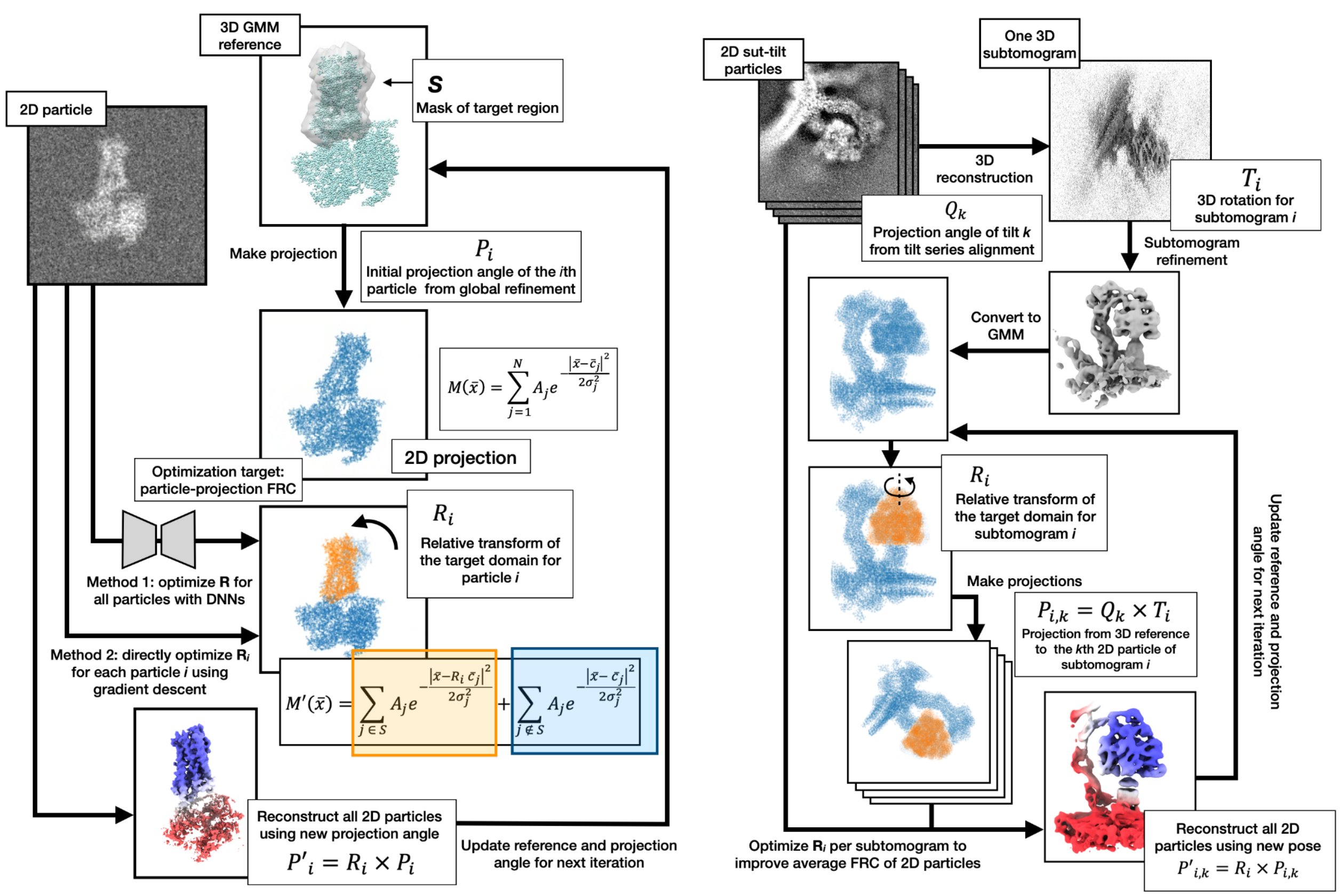

**Figure S1.** Workflow of GMM-based focused refinement for single particle CryoEM (left) and CryoET (right).

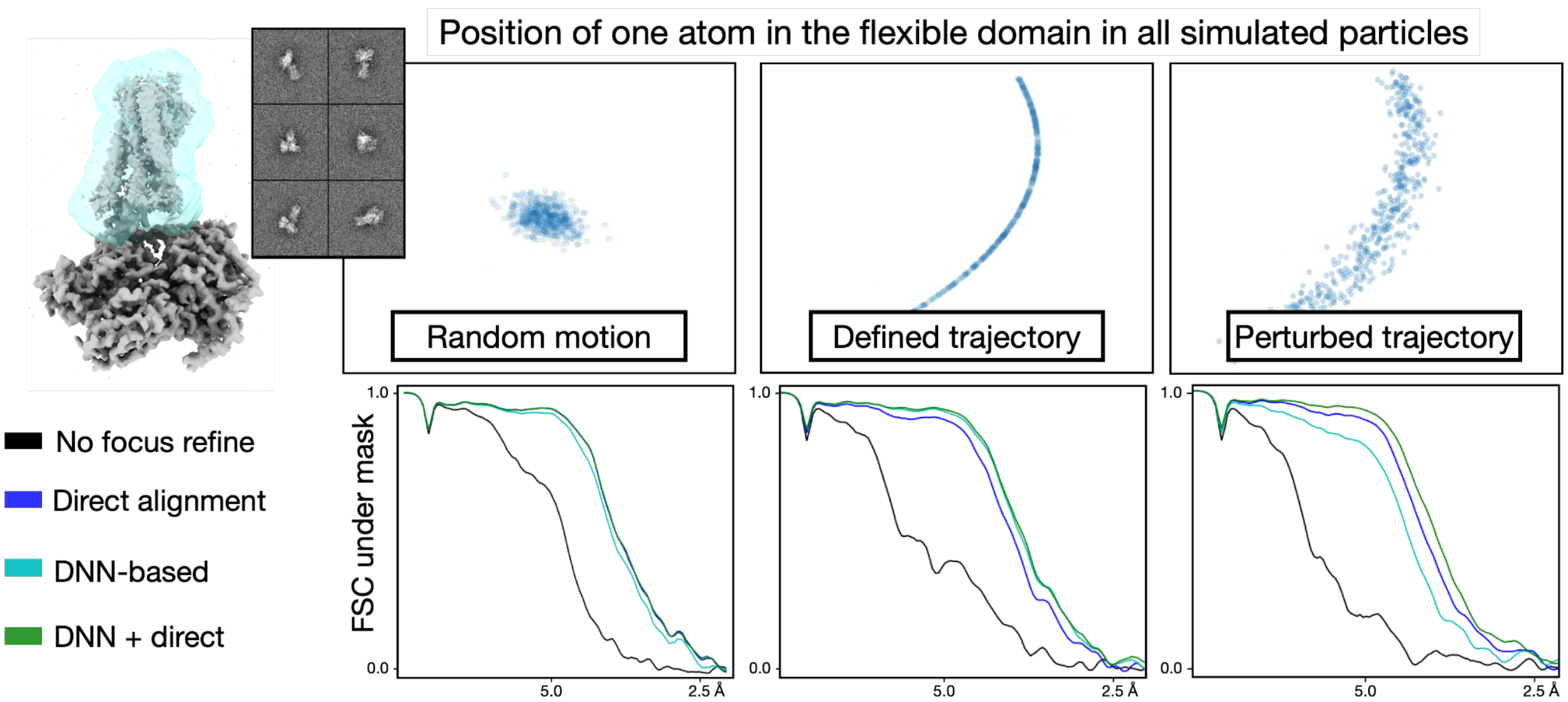

**Figure S2.** Performance of the two optimization methods on a simulated dataset. Different types of dynamics are simulated for the transmembrane domain (cyan mask). The results are evaluated by the FSC between focus-refined structure and the ground truth under the cyan mask.

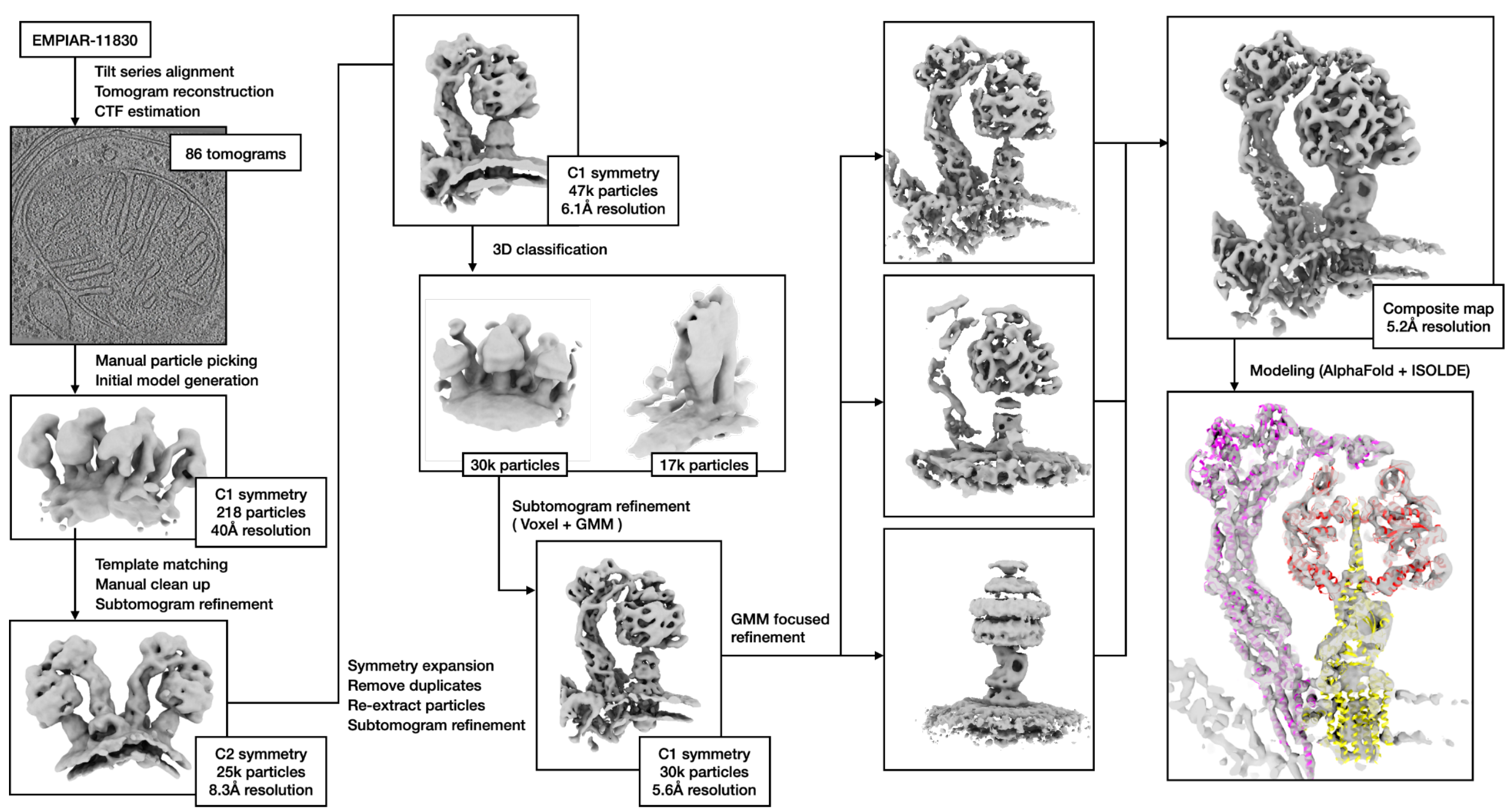

**Figure S3.** Data processing workflow for the mitochondrial ATP synthase CryoET dataset.